\documentclass[runningheads]{llncs}
\usepackage[T1]{fontenc}
\usepackage{graphicx}
\usepackage{subfigure}
\usepackage{amsmath}
\usepackage{amssymb}
\usepackage{tikz}
\usepackage{multirow}
\usepackage{siunitx}
\usepackage{standalone}
\usepackage{array}
\usepackage[pagebackref=true,breaklinks=true,letterpaper=true,colorlinks,bookmarks=false]{hyperref}
\usepackage{authblk} % Support for footnote style author/affiliation
\usepackage{longtable}% for long tables
\usepackage{floatrow}

\usepackage{capt-of}
\usepackage[]{graphicx}

\usepackage{booktabs}

\usepackage{capt-of}
\usepackage[]{graphicx}

\newfloatcommand{capbtabbox}{table}[][\FBwidth]

\begin{document}
\bibliographystyle{splncs04}

\title{Artifact Removal in Histopathology Images}
%
%\titlerunning{Abbreviated paper title}
% If the paper title is too long for the running head, you can set
% an abbreviated paper title here

\author{Cameron Dahan\inst{1}\and
Stergios Christodoulidis\inst{2}\and
Maria Vakalopoulou\inst{2}\and
Joseph Boyd\inst{2}}
%index{Boyd, Joseph}
%index{Villar, Irène}
%index{Mathieu, Marie-Christine}
%index{Deutsch, Eric}
%index{Paragios, Nikos}
%index{Vakalopoulou, Maria}
%index{Christodoulidis, Stergios}

\authorrunning{C. Dahan et al.}
% First names are abbreviated in the running head.
% If there are more than two authors, 'et al.' is used.

\institute{
CentraleSup\'elec, Universit\'e Paris-Saclay,
91190 Gif-sur-Yvette, France
\email{cameron.dahan@student-cs.fr} \and
MICS Laboratory, CentraleSup\'elec, Universit\'e Paris-Saclay,
91190 Gif-sur-Yvette, France
\email{firstname.lastname@centralesupelec.fr}}
\maketitle              % typeset the header of the contribution

\begin{abstract}
In the clinical setting of histopathology, whole-slide image (WSI) artifacts frequently arise, distorting regions of interest, and having a pernicious impact on WSI analysis. Image-to-image translation networks such as CycleGANs are in principle capable of learning an artifact removal function from unpaired data. However, we identify a surjection problem with artifact removal, and propose a weakly-supervised extension to CycleGAN to address this. We assemble a pan-cancer dataset comprising artifact and clean tiles from the TCGA database. Promising results highlight the soundness of our method.
\end{abstract}

\section{Introduction}

Due to their handling and large size, histopathology slides are susceptible to contaminants and other artifacts, leading to image artifacts in whole slide images (WSI). Common artifacts include pen marker, ink, blur, air bubbles, tissue folds, dust and filaments~\cite{smit2021quality}. Such artifacts create noise and outliers in WSI datasets, potentially undermining statistical analysis. An automatic system capable of localising and removing artifacts is therefore of great interest.

CycleGANs~\cite{zhu2017unpaired} are unpaired image-to-image translation models with several existing applications in histopathology, notably in stain normalisation~\cite{rana2018computational,bentaieb2017adversarial,de2018stain}, stain transfer~\cite{boyd2022region}, and cell segmentation~\cite{mahmood2019deep}. CycleGANs have been applied to histopathology artifacts before~\cite{ali2019ink}, but this was restricted to the relatively straightforward case of pen marker.

In Section \ref{sec:methods} we propose an extension to the CycleGAN framework aimed at accounting for the surjective nature of artifact removal. In Section \ref{sec:results} we demonstrate the improvement of our proposed method over baselines.

\section{Methods}
\label{sec:methods}

Our artifact removal system is based on the CycleGAN framework, for which we denote artifact and clean tiles as image domains $A$ and $B$. Accordingly, our baseline objective function is,

\begin{align}
\mathcal{L}_{base} &= \mathcal{L}_{GAN}(G_{AB}, D_B) + \lambda_{ABA}\cdot\mathcal{L}_{CYC}(G_{BA} \circ G_{AB}) + \lambda_{A}\cdot\mathcal{L}_{ID}(G_{BA}) \label{eq:cyc_loss} \\
&{}+  \mathcal{L}_{GAN}(G_{BA}, D_A) + \lambda_{BAB}\cdot\mathcal{L}_{CYC}(G_{AB} \circ G_{BA}) + \lambda_{B}\cdot\mathcal{L}_{ID}(G_{AB}), \notag
\end{align}

where generators $G_{AB}$ and $G_{BA}$ translate between the image domains, and $D_{B}$ and $D_{A}$ are their respective discriminators. The loss $\mathcal{L}_{GAN}$ represents a least squares GAN loss \cite{mao2017least}; $\mathcal{L}_{CYC}$ a cycle consistency loss; and $\mathcal{L}_{ID}$ an identity function loss. The system is optimised through a minmax optimisation process.
%\begin{equation}
%G_{AB}*, G_{BA}* = \arg\min_{G_{AB}, G_{BA}}\max_{D_{A}, D_{B}} \mathcal{L}.
%\end{equation}

The key observation of our method is that artifact removal is a \emph{surjective} operation; many artifact images may correspond with the same clean image. If truly clean, the image should not retain information about the artifact. This creates a difficult task for generator $G_{BA}$, which is required to reconstruct artifacts in $\lambda_{ABA}\cdot\mathcal{L}_{CYC}$ (Equation \ref{eq:cyc_loss}). This creates strong pressure for $G_{AB}$ to only partially clean or otherwise encode extraneous information within artifact images. This is clearly at odds with minimising the adversarial loss $\mathcal{L}_{GAN}(G_{AB}, D_B)$ (Equation \ref{eq:cyc_loss}). We refer to this as the surjection problem and hypothesise that this leads the CycleGAN to a poor local minimum. An initial idea is to simply set $\lambda_{ABA} = 0$, and to rely on the other half of the cycle to enforce consistency.

\subsection{Weakly supervised CycleGAN}

\begin{figure}
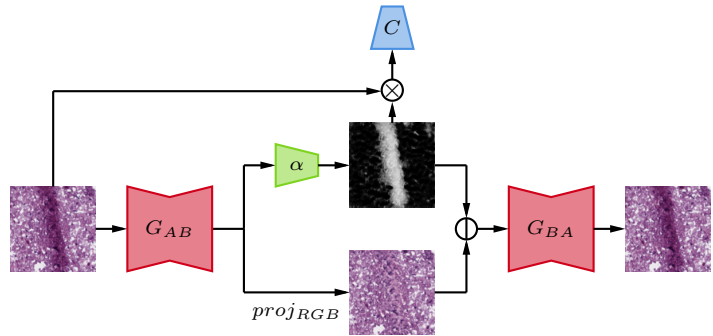

\centering
  \includestandalone[width=0.8\textwidth]{tikz/model}%     without .tex extension
  % or use \input{mytikz}
  \caption{Weakly-supervised CycleGAN for artifact$\to$clean$\to$artifact consistency.}
  \label{fig:diagram}
\end{figure}

To better address the problem of surjective artifact removal, we leverage the label of artifact tiles in two alternative models. In the first model, the label is used as a conditioning variable for both the $G_{BA}$ generator and the $D_A$ discriminator. For each, the condition is encoded as a one-hot tensor with $7$ class channels, where a channel is set to ones to indicate the class label. This conditioning is concatenated with the input image channel-wise. Thus, $G_{BA}$ can learn what class of artifact has been removed. We refer to this model as $\mathcal{M}_{cond}$.

In the second model we introduce an attention network, $\alpha$. This is a two-layer MLP applied to each pixel of the penultimate layer, $g_{AB}$, of generator $G_{AB}$, followed by a sigmoid activation function. The $A\to B\to A$ cycle-consistency is thus redefined to be,

\begin{equation}
G_{BA} \circ G_{AB} := G_{BA}\Big(concat\big(\alpha(g_{AB}), proj_{RGB}(g_{AB})\big)\Big)
\end{equation}

where $proj_{RGB}$ is a linear projection followed by a \emph{tanh} activation, and $concat$ is channel-wise concatenation. As such, $G_{BA}$ has an additional channel of input. Since the attention map is only available when computing the cycle-consistency loss, at all other times we substitute a dummy channel of Gaussian noise. This model is referred to as $\mathcal{M}_{attn}$.

To enhance $\mathcal{M}_{attn}$ we employ a similar trick as in~\cite{sahasrabudhe2020self}, where the attentions are combined with the image inputs by element-wise product, and then fed into an auxiliary convolutional network $C$, which classifies the artifact under a cross-entropy loss, $\mathcal{L}_{cls}$. The mask is regularised with smoothness $\mathcal{L}_{smooth}$ and sparsity $\mathcal{L}_{sparse}$ losses. This compels the attention model to produce masks that highlight the artifact regions only. Our hypothesis is that this will allow $G_{AB}$ to decouple the tissue from the artifact, and to offload all extraneous information onto the attention map. This model is referred to as $\mathcal{M}_{ws}$, and is depicted in Figure \ref{fig:diagram}. With $C$ guiding the attention map, the model becomes weakly-supervised, and the objective function becomes,

\begin{equation}
\mathcal{L}_{ws} = \mathcal{L}_{base} + \lambda_{cls}\cdot\mathcal{L}_{cls} + \lambda_{smooth}\cdot\mathcal{L}_{smooth} + \lambda_{sparse}\cdot\mathcal{L}_{sparse}
\end{equation}

\subsection{Dataset}

Our dataset consists of $28$ WSIs randomly selected from the TCGA database~\footnote{\url{https://www.cancer.gov/tcga}}, and thus spans a wide range of tissue types. $6556$ non-overlapping tiles of size $300\times300$px were manually extracted capturing typical artifact classes at both $40$x and $10$x magnification. For each artifact tile, a clean equivalent was extracted in close proximity, as in Supp. Figure 1, to control for sampling bias. Each artifact tile was labelled into one of seven classes: pen marker, ink, blur, air bubble, tissue fold, dust and filament, as shown in Supp. Figure 2.

\subsection{Model training and evaluation}

We train $\mathcal{L}_{base}$ with UNet and attention UNet~\cite{oktay2018attention} generators, which we denote $\mathcal{M}_{base}$ and $\mathcal{M}_{dpa}$ respectively. Apart from $\mathcal{M}_{\lambda_{ABA} = 0}$, the hyperparameters $\lambda_{ABA} = 5$, $\lambda_{BAB} = 5$, $\lambda_{A} = 5$, and $\lambda_{B} = 5$ are fixed for all models. For model $\mathcal{M}_{ws}$ we set $\lambda_{cls}=1$, $\lambda_{smooth}=1$, and $\lambda_{sparse}=0.1$.

We train all networks with the Adam optimiser~\cite{kingma2014adam}, with $lr = 0.001$, $\beta_1 = 0.5$, $\beta_2 = 0.999$, and an exponential learning rate decay of $\gamma = 0.9975$. All models were trained for 30 epochs with batch size 16. To regularise learning, we augment the data with random horizontal and vertical flips, followed by taking a random $256\times256$px crop and resizing to $128\times128$px. At test time the random crop is replaced by a center crop. We apply label smoothing to the discriminators and a weight decay of \num{1e-5} for all networks.

There is an inherent difficulty in evaluating unpaired datasets, as the targets are unknown. We therefore compare the distributional output of each model with the negative samples using Fréchet Inception Distance (FID) on both train (2620 samples) and test (655 samples) sets to evaluate the results (i.e. an $80$-$20$ split). Note that FID is conventionally evaluated over $50,000$ samples~\cite{heusel2017gans}, but this is not possible in our dataset.

\section{Results}
\label{sec:results}

% \begin{figure}
% \begin{floatrow}
% \ffigbox{%
%   \includegraphics[width=0.3\textwidth]{imgs/mosaic.png}
% }{%
%   \caption{A figure}%
% }
% \capbtabbox{%
% \begin{tabular}{ |l|c|c| } 
%  \hline
%  Model & Train FID & Test FID \\
%  \hline
% %  $bf = 16$ & 46.62 & 72.25 \\
% %  $bf = 16$ + dot-product attention & 50.04 & 75.87 \\ 
%  $\mathcal{M}_{base}$ & $45.09$ & 70.71 \\ 
%  $\mathcal{M}_{dpa}$ & 44.95 & 70.29 \\ 
%  \hline
%  $\mathcal{M}_{cond}$ & 42.62 & 67.93 \\
%  $\mathcal{M}_{\lambda_{ABA} = 0}$ & 42.38 & 67.72 \\ 
%  $\mathcal{M}_{attn}$ & 45.39 & 75.33 \\ 
%  $\mathcal{M}_{attn+clf}$ & $\mathbf{34.43}$ & $\mathbf{59.36}$  \\ 
%  \hline
% \end{tabular}
% }{%
% \label{table:results}
%   \caption{A table}%
% }
% \end{floatrow}
% \end{figure}

As shown in Table \ref{table:results}, the modified CycleGANs generally perform better, in particular $\mathcal{M}_{ws}$. The train FID scores are systematically lower than the test scores, but this does not seem to be a sign of overfitting, rather the effect of sample size. Controlling for sample size, our best model achieves a FID of $60.24$ on $655$ training samples and $59.36$ on the test data. We also note the baseline is slightly improved upon with dot product attention.

\begin{figure}[htp]
\begin{center}
\includegraphics[width=0.65\textwidth]{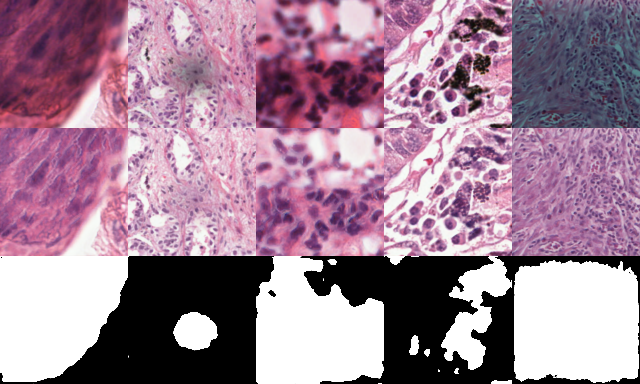}
\caption{Model inputs, outputs, and attention (top to bottom) for unseen test images.}
\label{fig:mosaic}
\end{center}
\end{figure}

Figure \ref{fig:mosaic} shows examples of artifact tiles from the test set, cleaned tile model outputs and corresponding attention maps. We note the localisation of artifact regions in the attention maps, which has been achieved from weak labels only. We include additional examples in Supp. Figure 3, as well as failure cases in Supp. Figure 4. Currently, the model struggles with opaque pen marker and underrepresented classes such as filament artifacts.

% \begin{table}
% \begin{center}
% \begin{tabular}{ |l|c|c| } 
%  \hline
%  Model & Train FID & Test FID \\
%  \hline
% %  $bf = 16$ & 46.62 & 72.25 \\
% %  $bf = 16$ + dot-product attention & 50.04 & 75.87 \\ 
%  $\mathcal{M}_{base}$ & $45.09$ & 70.71 \\ 
%  $\mathcal{M}_{dpa}$ & 44.95 & 70.29 \\ 
%  \hline
%  $\mathcal{M}_{cond}$ & 42.62 & 67.93 \\
%  $\mathcal{M}_{\lambda_{ABA} = 0}$ & 42.38 & 67.72 \\ 
%  $\mathcal{M}_{attn}$ & 45.39 & 75.33 \\ 
%  $\mathcal{M}_{attn+clf}$ & $\mathbf{34.43}$ & $\mathbf{59.36}$  \\ 
%  \hline
% \end{tabular}
% \end{center}
% \caption{FID results. Best results in bold.}
% \label{table:results}
% \end{table}

\begin{table}
\begin{center}
 \begin{tabular}{|l|p{1.5cm}|p{1.5cm}|p{1.5cm}|p{1.5cm}|p{1.5cm}|p{1.5cm}|} 
 \hline
  & $\mathcal{M}_{base}$ & $\mathcal{M}_{dpa}$ & $\mathcal{M}_{cond}$ & $\mathcal{M}_{\lambda_{ABA} = 0}$ & $\mathcal{M}_{attn}$ & $\mathcal{M}_{ws}$\\
 \hline
 Tr. FID & $45.09$ & 44.95 & 42.62 & 42.38 & 45.39 & $\mathbf{34.43}$ \\
 \hline
 Te. FID  & $70.71$ & 70.29 & 67.93 & 67.72 & 75.33 & $\mathbf{59.36}$ \\
 \hline
\end{tabular}
\end{center}
\caption{FID scores on train and test data. Best results in \textbf{bold}.}
\label{table:results}
\end{table}

\section{Conclusions}

In this paper we have identified a limitation of CycleGANs for the surjective task of artifact removal in histopathology images. We have presented a mechanism for incorporating weakly-supervised data into a CycleGAN, allowing it to decouple tissue from artifact, and improving over baselines in its ability to remove artifacts. An approximate artifact segmentation is a byproduct of the removal process. Future work could aim to expand the dataset, as currently some artifact classes are underrepresented.

%\acks{Acknowledgements should go at the end, before appendices and references. You can uncomment this for the camera-ready version on paper acceptance.}

%\bibliographystyle{plain}
\bibliography{acml22}

% \appendix

% \section{First Appendix}\label{apd:first}

% \begin{figure}
% \centering     %%% not \center
% \subfigure[Air Bubble]{\label{fig:a}\includegraphics[width=0.2\textwidth]{imgs/bubble.png}}
% \subfigure[Dust]{\label{fig:b}\includegraphics[width=0.2\textwidth]{imgs/dust.png}}
% \subfigure[Filament]{\label{fig:b}\includegraphics[width=0.2\textwidth]{imgs/filament.png}}
% \subfigure[Out of focus]{\label{fig:b}\includegraphics[width=0.2\textwidth]{imgs/focus.png}}
% \subfigure[Tissue fold]{\label{fig:b}\includegraphics[width=0.2\textwidth]{imgs/fold.png}}
% \subfigure[Ink]{\label{fig:b}\includegraphics[width=0.2\textwidth]{imgs/ink.png}}
% \subfigure[Pen marker]{\label{fig:b}\includegraphics[width=0.2\textwidth]{imgs/marker.png}}
% \caption{The seven categories of image artifact studied.}
% \label{fig:artifacts}
% \end{figure}

% \begin{figure}[htp]
% \begin{center}
% \includegraphics[width=0.8\textwidth]{imgs/annotations.png}
% \caption{Annotated region.}\label{fig:annotations}
% \end{center}
% \end{figure}

% \section{Second Appendix}\label{apd:second}

% This is the second appendix.

\appendix

\section{Supplementary Figures}\label{apd:first}

\begin{figure}[h]
\centering     %%% not \center
\subfigure[Air Bubble]{\label{fig:a}\includegraphics[width=0.2\textwidth]{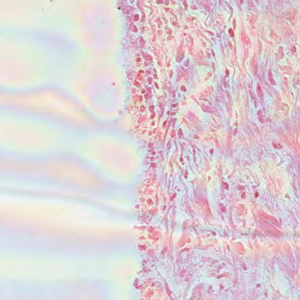}}
\subfigure[Dust]{\label{fig:b}\includegraphics[width=0.2\textwidth]{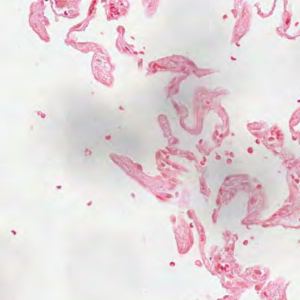}}
\subfigure[Filament]{\label{fig:b}\includegraphics[width=0.2\textwidth]{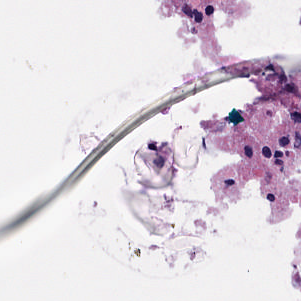}}
\subfigure[Out of focus]{\label{fig:b}\includegraphics[width=0.2\textwidth]{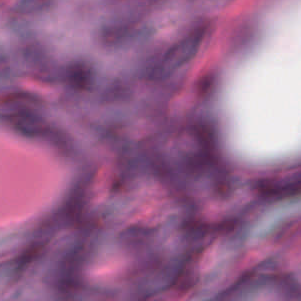}}
\subfigure[Tissue fold]{\label{fig:b}\includegraphics[width=0.2\textwidth]{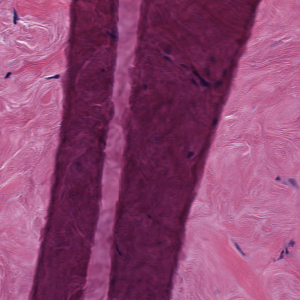}}
\subfigure[Ink]{\label{fig:b}\includegraphics[width=0.2\textwidth]{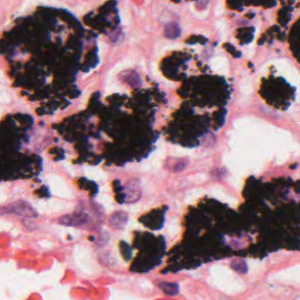}}
\subfigure[Pen marker]{\label{fig:b}\includegraphics[width=0.2\textwidth]{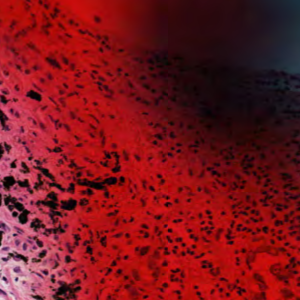}}
\caption{The seven categories of image artifact studied.}
\label{fig:artifacts}
\end{figure}

\begin{figure}[h]
\begin{center}
\includegraphics[width=0.7\textwidth]{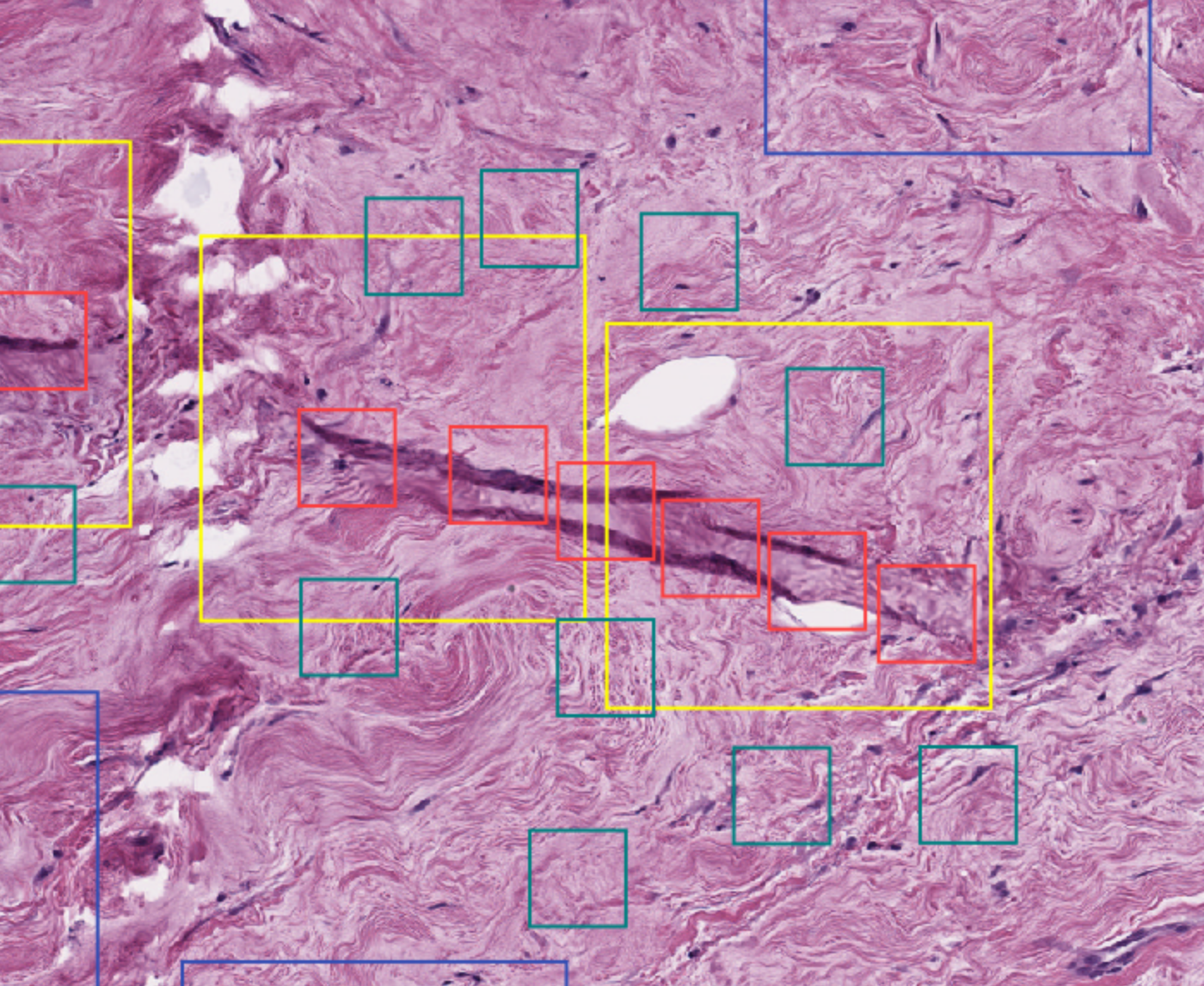}
\caption{Annotated tissue region comprising positive and negative tiles at two scales.}
\label{fig:annotations}
\end{center}
\end{figure}

% \section{Second Appendix}\label{apd:second}

% This is the second appendix.

\begin{figure}[htp]
\begin{center}
\includegraphics[width=0.75\textwidth]{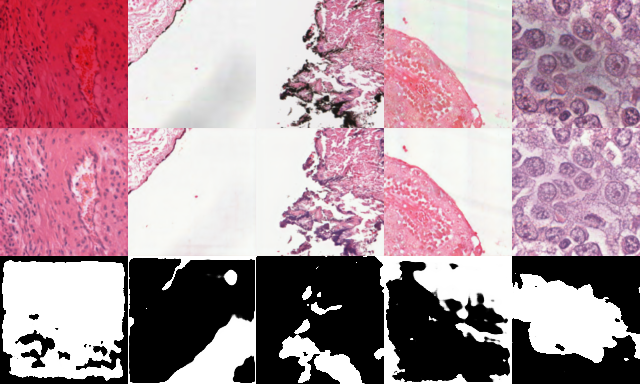}
\caption{Artifact removal samples.}
\label{fig:mosaic_success}
\end{center}
\end{figure}

\begin{figure}[htp]
\begin{center}
\includegraphics[width=0.75\textwidth]{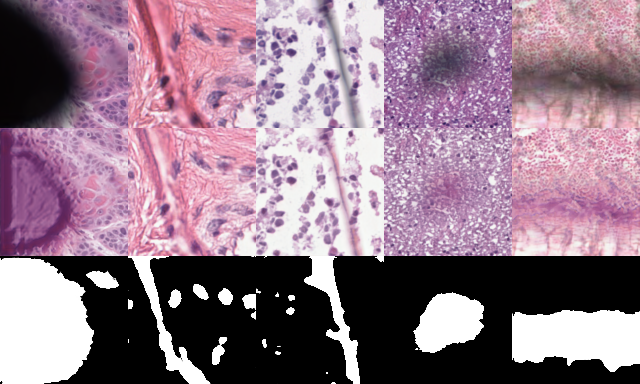}
\caption{Artifact removal failure cases.}
\label{fig:mosaic_failure}
\end{center}
\end{figure}

\end{document}